# Detection of cellular micromotion by advanced signal processing

Stephan Rinner[1,2,‡], Alberto Trentino[1,2,‡], Heike Url[4], M. Florian G. Burger[1,2], Julian von Lautz[3], Bernhard Wolfrum[4] and Friedemann Reinhard[1,2*]

[1] Technische Universität München, Walter Schottky Institut and Physik-Department, Garching, 85748, Germany
[2] Nanosystems Initiative Munich (NIM), Schellingstraße 4, 80799 München, Germany
[3] Landshuter Allee 27, 80637 München, Germany
[4] Technische Universität München, Munich School of Bioengineering, Department of Electrical and Computer Engineering, Boltzmannstraße 11, 85748 Garching, Germany
*friedemann.reinhard@wsi.tum.de
‡these authors contributed equally to this work

## Abstract

Cellular micromotion – a tiny movement of cell membranes on the nm-µm scale – has been proposed as a pathway for inter-cellular signal transduction and as a label-free proxy signal to neural activity. Here we harness several recent approaches of signal processing to detect such micromotion in video recordings of unlabeled cells. Our survey includes spectral filtering of the video signal, matched filtering, as well as 1D and 3D convolutional neural networks acting on pixel-wise time-domain data and a whole recording respectively.

## Introduction

There is considerable evidence that the intrinsic mechanical and optical properties of cells change slightly upon firing of an action potential. Signatures of such changes have been reported as early as 1949 [1] and have since been studied in a variety of channels. Action potentials slightly alter the birefringence of cell membranes, on a relative level of 10-100 ppm on a single cell [2–5]. This is plausibly explained by a Kerr effect induced by molecular alignment in the electric field, or changes in membrane thickness. Similar changes occur in light scattering, with a level of change of 1-1000 ppm for a single cell [2,6]. Less directly correlated with electrical activity, they are presumably linked to motion or swelling of cells. This 'intrinsic optical signal' [7–10] has been widely employed for the study of networks of neurons, both in a cell culture and in vivo in the retina [11]. Similar changes of transmission or reflection in near-infrared imaging of a living brain [12–14] have been controversially reported as a 'fast intrinsic signal' [15,16]. On the microscopic level, nanometer-scale motion of the cell membrane in response to an action potential has been observed by fiber-optical and piezoelectric sensors [17], atomic force microscopy [18] and optical interferometry [19–24]. More recently, such motion has been detected in non-interferometric microscope videos by image processing [25,26], which is the technique we intend to advance with the present work.

Intrinsic changes in optical or mechanical properties are of interest for two reasons. First, mechanical motion of cell membranes can be involved in cellular communication, driving for instance synchronization of heart muscle cells [27]. Second, intrinsic signals could provide access to neural activity. In contrast to existing fluorescent indicators [28], a method based on intrinsic signals would be label-free. It would not require genetic engineering and would not suffer from toxicity and photobleaching.

Previous studies have detected and quantified membrane micromotion by very simple schemes, such as manual tracking or subtraction of a static background image. The past decade has seen the emergence of numerous novel approaches to highlight small temporal changes in time series data, detecting for instance gravitational waves in interferometer signals [29] and invisible motion in real-life videos [30]. In the present work, we will study whether these tools can improve detection of cellular micromotion in video recordings of living cells. We focus our study on three of the most common approaches: spectral filtering, matched filtering, and convolutional neural networks (CNNs).

Spectral filtering has been a long-standing standard technique in the audio domain, where it is known as "equalizing". A time-domain signal is Fourier-transformed into the frequency domain, multiplied by a filter function that highlights or suppresses specific frequency bands, and subsequently transformed back into the time domain. It is equally applicable to video recordings [30], where it can detect and amplify otherwise invisible changes, such as the slight variation of skin color induced by blood circulation during a human heartbeat. Similarly, it should be able to detect micromotion of cells.

Matched filtering can be understood as an extension of spectral filtering. Here, the filter applied in the frequency domain is the Fourier transform of an ideal template signal. Employing this transform as a filter has a convenient interpretation in the time domain: it is a deconvolution of the signal with the template, i.e. a search for occurrence of the template in an unknown time series. Originally developed for radar processing [31], the technique has found ubiquitous applications. It is for instance employed to detect and count subthreshold events in gravitational wave detectors [29]. It has already been applied to the detection of mechanical deformation in videos [32] and should equally be applicable to cellular micromotion. It does, however, require a priori knowledge of an "ideal" template signal.

This drawback is overcome by neural networks, which can autonomously learn complex patterns and detect their occurrence in time-series data, images or video recordings. We focus on "convolutional neural networks" (CNNs), a widely employed subclass of networks that can be understood as an extension of matched filtering. In a CNN, an unknown input signal is repeatedly convolved with a set of simple patterns ("filters") and subjected to a nonlinear "activation function". Repetition of this process greatly enlarges the range of patterns that can be detected, so that the technique can detect patterns even if they deviate from some fixed ideal signal. The pattern vocabulary of the network is learned in a training procedure, in our case in a "supervised" fashion where the network is optimized to detect known occurrences of a pattern in a separate training dataset. During training, the filters are continuously adapted to improve the detection fidelity. The result is equivalent to repeated application of matched filtering with 'learned' filters, interleaved with nonlinear elements. CNNs have been implemented for datasets of various dimensions. One-dimensional CNNs have found use in time-series processing, most prominently speech recognition [33], two-dimensional CNNs in image recognition [34] and three-dimensional networks in video analysis [35].

## Methods

All techniques are trained and benchmarked on a dataset recorded as displayed in Fig. 1. A sample of HL-1 cardiac cells (originating from the Claycomb lab[36]) is recorded in a homebuilt dual-channel video microscope (Fig. 1 (a)). These cells fire spontaneous action potentials every few seconds, which are accompanied by micromotion on a wide range of motion amplitudes (see below). Hence, they provide a convenient testbed to evaluate signal processing. One channel of the microscope performs imaging in transmission mode under strong brightfield illumination (Fig. 1 (b)). We choose a camera with high frame rate (50 fps) and full-well-capacity ($10^5$ electrons, obtained by averaging over 10 consecutive frames of a 500 fps recording) to reduce photon shot noise and thus enhance sensitivity to small changes in the image. Illumination is polarized, and detection is slightly polarization-selective, in order to be sensitive to small changes in cellular birefringence, although we did not find evidence for such a signal in the final data. A second channel records fluorescence of a $Ca^{2+}$-active dye (Cal-520) staining the cells (Fig. 1 (b)). $Ca^{2+}$ transients correlating with the generation of action potentials are directly visible in this channel as spikes of fluorescence and serve as a "ground truth" signal for training of the signal processing algorithms. We will identify these transients with individual action potentials [37], although our setup is lacking electrophysiological means to strictly prove this connection. All following analysis is based on a video recording of 2:30 minutes length. This dataset is divided into two parts for validation (frames 1 to 3072) and training (frames 3073 to 10570) of the processing schemes.

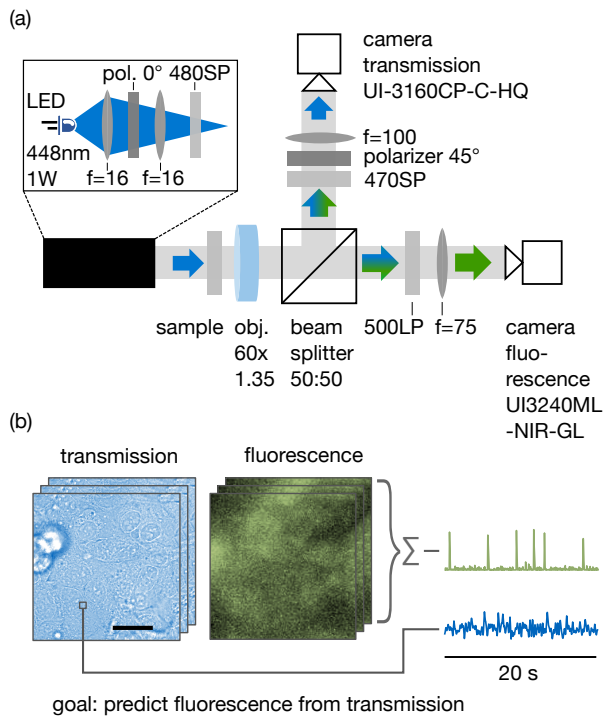

**Figure 1 - Experimental setup and data. (a) Experimental setup. A correlative video microscope records a sample of cells in two channels: light transmitted under brightfield illumination and fluorescence of a Ca-active staining. LP: long pass, SP: short pass, pol.: polarizer. (b) Resulting data. A region of several cells is visible in the transmission channel (Scale bar: 20 μm.). The same region displays spikes of Ca activity in the fluorescence channel. The fluorescence intensity of the whole region is summed to a time trace, which is employed as ground truth for supervised learning.**

All signal processing schemes under study are tasked with the same challenge: to predict fluorescence activity from the transmission signal (Fig. 2 (a)). The algorithms employed are summarized in Fig. 2. We implement spectral filtering by applying a temporal bandpass filter to every pixel of the recording (Fig. 2 (b)). Changes in transmission intensity, as they are produced by motion of the cells, will pass this filter, while both the static background image and fast fluctuating noise are suppressed. The filter parameters have been manually tuned to match the timescale of cellular motion, resulting in a 3 dB passband from 2 Hz to 15 Hz. Processing by this filter serves as an initial stage in all other algorithms, for reasons to be discussed below.

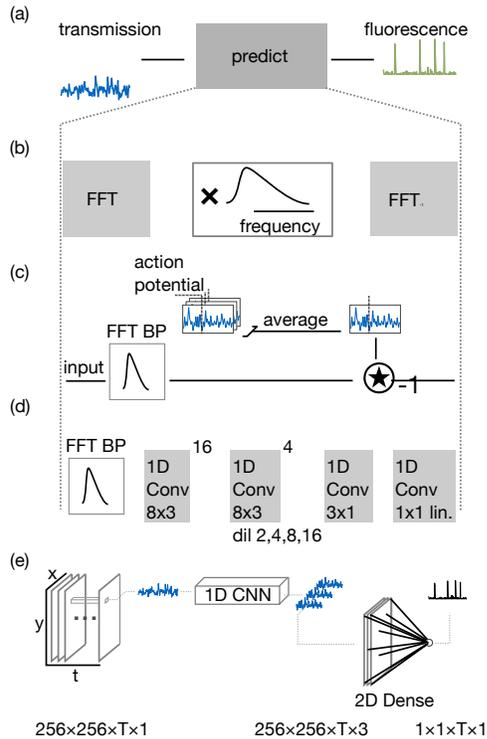

**Figure 2 - Signal processing schemes to detect cellular micromotion (a) concept: signal processing is employed to predict fluorescence from micromotion cues in transmission data. (b-d) present schemes processing time-domain data from a single pixel, (e) presents 3D neural networks processing a whole recording in the temporal and spatial domains.**

Matched filtering is implemented as an additional stage of processing as displayed in Fig. 2 (c). We generate a pixel-wise template signal by averaging over all 51 action potentials obtained from the training dataset, aligned by triggering on spikes in the fluorescence channel. These template signals are generated in a pixel-wise fashion so that this scheme should capture arbitrary signal shapes, such as upward and downward excursions of the video signal that might occur on two sides of a moving cell membrane. The prediction is computed by pixel-wise deconvolution of the transmission signal with the template.

One-dimensional convolutional neural networks equally serve as an additional stage of processing downstream of spectral filtering (Fig. 2 (d)). We stack 20 convolutional layers with eight filters of three frames width each, the last four of which are dilated with an exponentially increasing rate to capture features extending over long timescales [38] (see Table 1 for an exact description). The number of features (filters) is subsequently condensed to three and finally to only one, which provides the output prediction. Padding in the convolutional layers ensures that the temporal length of the data is preserved throughout the entire network, so that the output is a time series of the same size as the input. The network can therefore be equally understood as a cross-encoder architecture translating transmitted light into a prediction of fluorescence. The choice of this network architecture has been motivated by its simplicity, and by encouraging reports on conceptually similar fully convolutional neural networks that have proven successful in classification of electrocardiogram data [39]. Recurrent neural networks, frequently employed in speech recognition, would be another natural choice. They have not been pursued further in this study, because of the widely held belief that they are more difficult to train [39].

The network is trained to predict the fluorescence intensity summed over the full frame from single-pixel input data. This is of course an impossible challenge for all those pixels which do not contain a trace of cell motion, such as background regions. However, we found that convergence of the network was stable despite this conceptual weakness. Networks have been initialized with Glorot-uniform weights and have been trained for 3 epochs at a learning rate of $10^{-3}$ without weight decay, using the training data described above and mean-square-error as a loss function. Training has been carried out by the Adam algorithm, which locally adapts the learning rate for every weight and time step. We tried to train on raw data that had not undergone spectral filtering, but did not achieve convergence in this case. Due to the simple architecture, optimization of the hyperparameters is mainly limited to varying the number of layers. Here we found that performance generally improves with network depth, despite the fact that the patterns to be detected are relatively simple. This observation is in line with previous reports on time series classification by CNNs [40]. We also found dilations to provide a significant gain in performance. This suggests that the network successfully captures temporally extended patterns. We therefore did not venture into adding and optimizing pooling layers, which are frequently used for the same goal [29].

As the most elaborate approach we apply three-dimensional neural networks (Fig. 2 (e)). They operate on a 256 frames long section of the video recording, predicting a 256 frames long fluorescence trace. For 3D networks, limited training data is a major challenge, which we address by two means. First, we employ transfer learning, by reusing the one-dimensional network (Fig. 2 (d)) as an initial stage of processing. We terminate this 1D processing before the final layer, providing three output features for every pixel and timestep (see Table 2 for an exact description) that serve as input to a newly trained final layer. Second, we restrict ourselves to region-specific networks that cannot operate on video recordings of another set of cells. This simplifies processing of the spatial degrees of freedom, which is implemented by a simple dense layer connecting every feature in every pixel to one final output neuron computing the prediction. As in the case of matched filters, this stage can learn pixel-specific patterns, such as upward and downward excursions of light intensity during an action potential. The network was trained for 60 epochs by Adam with a learning rate of $8 \cdot 10^{-7}$ without weight decay. The 1D processing layers were initialized with the 1D network described above (Fig. 2 (d)), but were not held fixed during training. The final layer was initialized with the three final-layer weights of the 1D network divided by the number of pixels.

All networks have been defined in Keras using the Tensorflow backend, and all training has been performed on Google Cloud.

## Results

The performance of all one-dimensional processing schemes (Fig 2 (b)-(d)) is compared in Fig. 3. We employ cross-correlation as a score to benchmark how well the predicted fluorescence time trace of a specific pixel matches the global fluorescence intensity. Specifically, we normalize prediction and global fluorescence in a first step by subtracting the temporal mean and dividing by temporal standard deviation, since otherwise correct prediction of a constant background signal would be rated more important than correct prediction of spike signals. We then compute the full cross-correlation of both signals and employ the maximum as a figure of merit. This approach assigns a high score to predictions that are correct except for a constant temporal shift, which would be less detrimental to a human user than errors in detection of spikes.

All processing schemes succeed in detecting micromotion correlating with Calcium activity, with varying levels of success. The self-learning matched filter (Fig. 3 (c)) does not improve performance over mere spectral filtering (Fig. 3 (b)). This is likely caused by noise in the data. Since the template is generated on the single-pixel level, it will contain a higher level of noise than a smooth handcrafted filter or the filters of a neural network that have been trained on a much larger amount of data (all pixels). This error is inherent to the training technique rather than the technique of matched filtering by itself. With a better template, performance would likely improve. 1D neural networks offer a clear gain in performance (Fig. 3 (d)), even in region where signals are weak (lower right), consistent with the intuition that training on all pixels reduces noise artefacts.

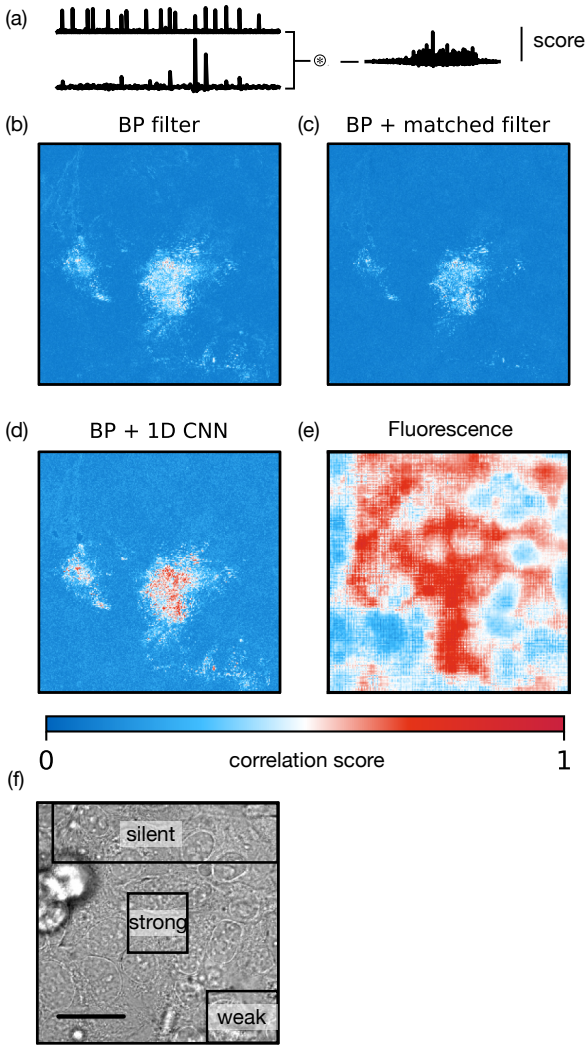

**Figure 3 - Performance of time-domain signal processing. (a) definition of correlation score. Predicted fluorescence on the single-pixel level is correlated with observed fluorescence summed over the full region. The maximum correlation is used as a score to assess accuracy of the prediction. (b-e) pixel-wise maps of correlation score for (b) band-pass filtering, (c) matched-filtering and (d) processing by a 1D CNN as defined in Fig. 2. (e) fluorescence activity (ground truth) (f) still frame from transmission channel. Labels denote regions of interest displaying strong motion, weak motion and no visible motion that will serve as test cases in the following analysis (Fig. 4). Scale bar: 20 μm.**

Fig. 4 analyzes performance in terms of three regions of interest (Fig. 3 (f)). One region (center) contains cells that display strong beating motion of 600 nm amplitude (measured by visually tracking membrane motion). A second region (lower right) contains motion on a weak level that is barely noticeable to the naked eye, suggesting an amplitude of significantly less than one pixel (i.e. ≪ 300 nm) or motion in an out-of-focus plane.  A third "silent" region does not contain any visible motion (upper). Fig. 4 shows the 1D predictions, binned over these regions of interest (ROI), as well as the prediction of the 3D network that has been separately trained and tested on each ROI. As in the correlation analysis (Fig. 3), the
pg. 7

difference of performance is most striking in the region of weak beating, where neural networks deliver a clear gain in performance. 3D networks perform marginally better than one-dimensional approaches. No approach is able to reveal a meaningful signal in the silent region.

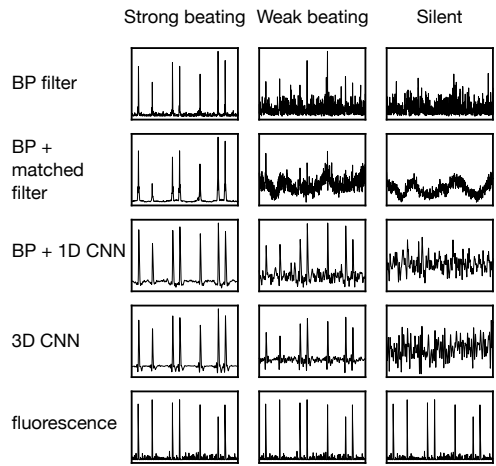

**Figure 4 - Performance of all considered schemes. Signals of 1D predictions (upper three lines) have been summed over the regions of interest marked in Fig. 3. The output of filtering approaches (upper two lines) has been squared to produce unipolar data comparable to fluorescence. All approaches manage to correctly predict fluorescence in the strong beating region. Performance varies in the weak beating region, where neural networks yield a clear gain in accuracy. No approach is able to reveal a meaningful signal in the silent region. Length of the recording is 20s.**

We finally analyze the dense layers of the 3D neural networks by visualizing their weights (Fig. 5), reasoning that these encode "attention" to specific pixels, where micromotion leaves a pronounced imprint. Strong weights are placed within a large homogeneous part of the strong beating region (Fig. 5 (a)), presumably a single cell. The boundary of the region is not clearly visible in the raw microscope image, which might be due to the high confluency in this region, or due to the cell sitting in a different layer than the image plane. Weights are placed very differently in the weak beating region (Fig. 5 (b)), where attention is mostly drawn to a small area at the border of a cell, presumably because small motions produce the most prominent signal change in this place. Interestingly, a similar behavior is observed in the silent region (Fig. 5 (c)). Most strong weights are placed on one membrane, even though the network does not detect a meaningful signal. This might be a sign that the network is overfitting to fluctuations rather than a real signal, since these are stronger at a membrane. It might equally hint towards existence of a small signal that could be revealed by further training on a more extensive dataset. Besides more extensive training, adding batch normalization and stronger dropout could be numerical options to improve generalization in these more challenging regions.

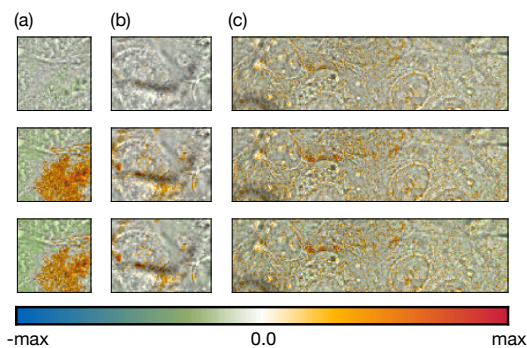

**Figure 5 - 3D neural networks. Weights of the fully connected layer ("2D Dense" in Fig. 2e), connecting three activity maps to the final output neuron. Weights are encoded in color and overlayed onto a still frame of the transmission video. The color scale is adjusted for each region; max: maximum weight occurring in all three layers of one region. (a) Strong beating region. Weights are placed on a confined region, presumably a single cell. (b) Weak beating region. Weights are predominantly placed on the border of one cell, where intensity is most heavily affected by membrane motion. (c) Silent region. While no meaningful prediction is obtained, the network does place weights preferentially on the border of one cell, hinting towards micromotion.**

## Discussion

In summary, recent schemes of signal processing can effectively detect and amplify small fluctuations in video recordings, revealing tiny visual cues such as micromotion of cells correlating with Calcium activity (and hence, presumably, with action potentials). Neural networks provide a clear gain in performance and flexibility over simpler schemes. They can be efficiently implemented, even with limited training data, if their architecture is sufficiently simplified. We achieve this simplification by one-dimensional processing of single-pixel time series, and by constructing more complex network from these one-dimensional ancestors via transfer learning. While the schemes of this study successfully predict Calcium activity from cellular micromotion for some cells, none of them is truly generic, i.e. able to provide a reliable prediction for any given cell. This is evidenced by performance in the "silent" region where no activity is detected, despite a clear signal of Calcium activity in fluorescence imaging. We foresee several experimental levers to overcome this limitation. Future experiments could improve detection of micromotion by recording with higher (kHz) framerate. This promises to reveal signals on a millisecond timescale, which would be the timescale of electric activity and some previously reported micromotion [20,21,41]. Illumination by coherent light could further enhance small motions [42] and thus reduce photon load of the cells, and more advanced microscopy schemes such as phase imaging or differential interference contrast could be employed to enhance the signal. The neural network architecture could be improved as well. Recurrent neural networks could be used for 1D processing, although the success of 1D-CNNs renders this direction less interesting. Instead, a major goal of our future work will be pushing 3D convolutional networks to a truly generic technique that is no longer restricted to one region of cells. This could be achieved by more complex architectures, such as the use of spatial 2D convolutions within the network rather than a single dense connection in the final layer. Recording orders of magnitude more training data (hours instead of minutes) will be another straightforward experimental improvement that likely needs to be addressed to successfully work with these architectures.

The ultimate performance of all schemes can be estimated from the photon shot noise limit of the acquisition chain. This limit predicts the minimum relative fluctuation of intensity that can be detected in a signal binned over $N_{\text{frames}}$ frames in the temporal dimension, and a region of $N_{\text{px}}$ pixels in the spatial dimension:

$$\frac{\sigma_I}{I} = \frac{1}{\sqrt{N_{\text{frames}} N_{\text{px}} N_{\text{FWC}}}}$$

Here, $N_{\text{FWC}}$ is the full-well capacity of the camera, typically on the order of $10^4$ photons. With current hardware, operating at 10 kHz frame rate, $N_{\text{frames}} = 10$ frames could be acquired over a 1 ms long action potential. Tracking the motion of a cell border extending over $N_{\text{px}} \approx 10^2$ pixels could thus reveal intensity fluctuations of $\frac{\sigma_I}{I} \approx 3 \cdot 10^{-4}$. Assuming a pixel size of 300 nm, and a modulation of the light intensity by the membrane of 10%, this would correspond to motion of $\approx 1$ nm. Label-free detection of single action potentials, with a reported amplitude of several nanometers[24], appears well within reach for future experiments.

## Data availability

All data and source code is available from the corresponding author upon request.

## References

1. Hill, D. K. & Keynes, R. D. Opacity changes in stimulated nerve. *The Journal of Physiology* **108**, 278–281 (1949).

2. Cohen, L. B., Keynes, R. D. & Hille, B. Light Scattering and Birefringence Changes during Nerve Activity. *Nature* **218**, 438–441 (1968).

3. Cohen, L. B., Hille, B. & Keynes, R. D. Changes in axon birefringence during the action potential. *The Journal of Physiology* **211**, 495–515 (1970).

4. Badreddine, A. H., Jordan, T. & Bigio, I. J. Real-time imaging of action potentials in nerves using changes in birefringence. *Biomed. Opt. Express, BOE* **7**, 1966–1973 (2016).

5. Foust, A. J. & Rector, D. M. Optically teasing apart neural swelling and depolarization. *Neuroscience* **145**, 887–899 (2007).

6. Stepnoski, R. A. *et al.* Noninvasive detection of changes in membrane potential in cultured neurons by light scattering. *PNAS* **88**, 9382–9386 (1991).

7. Grinvald, A., Lieke, E., Frostig, R. D., Gilbert, C. D. & Wiesel, T. N. Functional architecture of cortex revealed by optical imaging of intrinsic signals. *Nature* **324**, 361–364 (1986).
8. MacVicar, B. A. & Hochman, D. Imaging of synaptically evoked intrinsic optical signals in hippocampal slices. *J. Neurosci.* **11**, 1458–1469 (1991).
9. Andrew, R. D., Jarvis, C. R. & Obeidat, A. S. Potential Sources of Intrinsic Optical Signals Imaged in Live Brain Slices. *Methods* **18**, 185–196 (1999).
10. Yao, X.-C. Intrinsic optical signal imaging of retinal activation. *Jpn J Ophthalmol* **53**, 327–333 (2009).
11. Wang, B., Lu, Y. & Yao, X. In vivo optical coherence tomography of stimulus-evoked intrinsic optical signals in mouse retinas. *JBO* **21**, 096010 (2016).
12. Villringer, A., Planck, J., Hock, C., Schleinkofer, L. & Dirnagl, U. Near infrared spectroscopy (NIRS): A new tool to study hemodynamic changes during activation of brain function in human adults. *Neuroscience Letters* **154**, 101–104 (1993).
13. Cui, X., Bray, S., Bryant, D. M., Glover, G. H. & Reiss, A. L. A quantitative comparison of NIRS and fMRI across multiple cognitive tasks. *NeuroImage* **54**, 2808–2821 (2011).
14. Fazli, S. *et al.* Enhanced performance by a hybrid NIRS–EEG brain computer interface. *NeuroImage* **59**, 519–529 (2012).
15. Steinbrink, J., Kempf, F. C. D., Villringer, A. & Obrig, H. The fast optical signal—Robust or elusive when non-invasively measured in the human adult? *NeuroImage* **26**, 996–1008 (2005).
16. Morren, G. *et al.* Detection of fast neuronal signals in the motor cortex from functional near infrared spectroscopy measurements using independent component analysis. *Med. Biol. Eng. Comput.* **42**, 92–99 (2004).
17. Iwasa, K., Tasaki, I. & Gibbons, R. C. Swelling of nerve fibers associated with action potentials. *Science* **210**, 338–339 (1980).
18. Zhang, P.-C., Keleshian, A. M. & Sachs, F. Voltage-induced membrane movement. *Nature* **413**, 428–432 (2001).

19. Hill, B. C., Schubert, E. D., Nokes, M. A. & Michelson, R. P. Laser interferometer measurement of changes in crayfish axon diameter concurrent with action potential. *Science* **196**, 426–428 (1977).

20. Akkin, T., Davé, D. P., Milner, T. E. & Iii, H. G. R. Detection of neural activity using phase-sensitive optical low-coherence reflectometry. *Opt. Express, OE* **12**, 2377–2386 (2004).

21. Fang-Yen, C., Chu, M. C., Seung, H. S., Dasari, R. R. & Feld, M. S. Noncontact measurement of nerve displacement during action potential with a dual-beam low-coherence interferometer. *Opt. Lett., OL* **29**, 2028–2030 (2004).

22. Oh, S. *et al.* Label-Free Imaging of Membrane Potential Using Membrane Electromotility. *Biophysical Journal* **103**, 11–18 (2012).

23. Batabyal, S. *et al.* Label-free optical detection of action potential in mammalian neurons. *Biomed. Opt. Express, BOE* **8**, 3700–3713 (2017).

24. Ling, T. *et al.* Full-field interferometric imaging of propagating action potentials. *Light: Science & Applications* **7**, 1–11 (2018).

25. Fields, R. D. Signaling by Neuronal Swelling. *Sci. Signal.* **4**, tr1–tr1 (2011).

26. Yang, Y. *et al.* Imaging Action Potential in Single Mammalian Neurons by Tracking the Accompanying Sub-Nanometer Mechanical Motion. *ACS Nano* **12**, 4186–4193 (2018).

27. Tang, X., Bajaj, P., Bashir, R. & Saif, T. A. How far cardiac cells can see each other mechanically. *Soft Matter* **7**, 6151–6158 (2011).

28. Abdelfattah, A. S. *et al.* Bright and photostable chemigenetic indicators for extended in vivo voltage imaging. *Science* **365**, 699–704 (2019).

29. Gabbard, H., Williams, M., Hayes, F. & Messenger, C. Matching Matched Filtering with Deep Networks for Gravitational-Wave Astronomy. *Phys. Rev. Lett.* **120**, 141103 (2018).

30. Wu, H.-Y. *et al.* Eulerian Video Magnification for Revealing Subtle Changes in the World. *ACM Trans. Graph. (Proceedings SIGGRAPH 2012)* **31**, (2012).

31. North, D. O. An Analysis of the factors which determine signal/noise discrimination in pulsed-carrier systems. *Proceedings of the IEEE* **51**, 1016–1027 (1963).

32. Dansereau, D. G., Singh, S. P. N. & Leitner, J. Interactive computational imaging for deformable object analysis. in *2016 IEEE International Conference on Robotics and Automation (ICRA)* 4914–4921 (2016). doi:10.1109/ICRA.2016.7487697.

33. Abdel-Hamid, O. *et al.* Convolutional Neural Networks for Speech Recognition. *IEEE/ACM Transactions on Audio, Speech, and Language Processing* **22**, 1533–1545 (2014).

34. Krizhevsky, A., Sutskever, I. & Hinton, G. E. ImageNet Classification with Deep Convolutional Neural Networks. in *Advances in Neural Information Processing Systems 25* (eds. Pereira, F., Burges, C. J. C., Bottou, L. & Weinberger, K. Q.) 1097–1105 (Curran Associates, Inc., 2012).

35. Ji, S., Xu, W., Yang, M. & Yu, K. 3D Convolutional Neural Networks for Human Action Recognition. *IEEE Transactions on Pattern Analysis and Machine Intelligence* **35**, 221–231 (2013).

36. Claycomb, W. C. *et al.* HL-1 cells: A cardiac muscle cell line that contracts and retains phenotypic characteristics of the adult cardiomyocyte. *PNAS* **95**, 2979–2984 (1998).

37. Prajapati, C., Pölönen, R.-P. & Aalto-Setälä, K. Simultaneous recordings of action potentials and calcium transients from human induced pluripotent stem cell derived cardiomyocytes. *Biology Open* **7**, (2018).

38. Yu, F. & Koltun, V. Multi-Scale Context Aggregation by Dilated Convolutions. *arXiv:1511.07122 [cs]* (2016).

39. Ismail Fawaz, H., Forestier, G., Weber, J., Idoumghar, L. & Muller, P.-A. Deep learning for time series classification: a review. *Data Min Knowl Disc* **33**, 917–963 (2019).

40. Dai, W., Dai, C., Qu, S., Li, J. & Das, S. Very deep convolutional neural networks for raw waveforms. in *2017 IEEE International Conference on Acoustics, Speech and Signal Processing (ICASSP)* 421–425 (2017). doi:10.1109/ICASSP.2017.7952190.

41. LaPorta, A. & Kleinfeld, D. Interferometric Detection of Action Potentials. *Cold Spring Harb Protoc* **2012**, pdb.ip068148 (2012).

42. Taylor, R. W. & Sandoghdar, V. Interferometric Scattering Microscopy: Seeing Single Nanoparticles and Molecules via Rayleigh Scattering. *Nano Lett.* **19**, 4827–4835 (2019).

## Competing interests (mandatory)

The authors declare no competing interests. This project has received support from the Deutsche Forschungsgemeinschaft (DFG, German Research Foundation) under the German Excellence Strategy (NIM) and Emmy Noether grant RE3606/1-1, as well as from Bayerisches Staatsministerium für Wirtschaft, Landesentwicklung und Energie (IUK542/002). The authors thank Kirstine Berg-Sørensen for helpful discussions.

## Acknowledgements (optional)

This project has received support from the Deutsche Forschungsgemeinschaft (DFG, German Research Foundation) under the German Excellence Strategy (NIM) and Emmy Noether grant RE3606/1-1, as well as from Bayerisches Staatsministerium für Wirtschaft, Landesentwicklung und Energie (IUK542/002). The authors thank Kirstine Berg-Sørensen for helpful discussions.

## Author contributions

F.R. conceived the experiment and supervised it with B.W., A.T. and S.R. built the experimental setup and performed the experiment together with H.U. S.R., F.B. and F.R. developed the data processing with input from J.v.L. F.R. wrote the paper. All authors reviewed the manuscript.

## Figure Legends

**Figure 1 - Experimental setup and data. (a) Experimental setup. A correlative video microscope records a sample of cells in two channels: light transmitted under brightfield illumination and fluorescence of a Ca-active staining. LP: long pass, SP: short pass, pol.: polarizer. (b) Resulting data. A region of several cells is visible in the transmission channel (Scale bar: 20 μm.). The same region displays spikes of Ca activity in the fluorescence channel. The fluorescence intensity of the whole region is summed to a time trace, which is employed as ground truth for supervised learning.**

**Figure 2 - Signal processing schemes to detect cellular micromotion (a) concept: signal processing is employed to predict fluorescence from micromotion cues in transmission data. (b-d) present schemes processing time-domain data from a single pixel, (e) presents 3D neural networks processing a whole recording in the temporal and spatial domains.**

**Figure 3 - Performance of time-domain signal processing. (a) definition of correlation score. Predicted fluorescence on the single-pixel level is correlated with observed fluorescence summed over the full region. The maximum correlation is used as a score to assess accuracy of the prediction. (b-e) pixel-wise maps of correlation score for (b) band-pass filtering, (c) matched-filtering and (d) processing by a 1D CNN as defined in Fig. 2. (e) fluorescence activity (ground truth) (f) still frame from transmission channel. Labels denote regions of interest displaying strong motion, weak motion and no visible motion that will serve as test cases in the following analysis (Fig. 4). Scale bar: 20 μm.**

**Figure 4 - Performance of all considered schemes. Signals of 1D predictions (upper three lines) have been summed over the regions of interest marked in Fig. 3. The output of filtering approaches (upper two lines) has been squared to produce unipolar data comparable to fluorescence. All**

**approaches manage to correctly predict fluorescence in the strong beating region. Performance varies in the weak beating region, where neural networks yield a clear gain in accuracy. No approach is able to reveal a meaningful signal in the silent region. Length of the recording is 20s.**

**Figure 5 - 3D neural networks. Weights of the fully connected layer ("2D Dense" in Fig. 2e), connecting three activity maps to the final output neuron. Weights are encoded in color and overlayed onto a still frame of the transmission video. The color scale is adjusted for each region; max: maximum weight occurring in all three layers of one region. (a) Strong beating region. Weights are placed on a confined region, presumably a single cell. (b) Weak beating region. Weights are predominantly placed on the border of one cell, where intensity is most heavily affected by membrane motion. (c) Silent region. While no meaningful prediction is obtained, the network does place weights preferentially on the border of one cell, hinting towards micromotion.**

Tables

**Table 1 - Time-Domain 1D Convolutional Network. The input signal (of length T) is passed through 16 1D convolutions (Conv1 … Conv16), followed by 4 à trous convolutions with increasing dilation rate. ReLU: rectified linear unit. The final network is available as Data File 1.**

|  | Input | Conv1 | … | Conv16 | Conv17 | Conv18 | Conv19 | Conv20 | Conv21 | Conv22 |
|---|---|---|---|---|---|---|---|---|---|---|
| output dim | T×1 | T×8 |  | T×8 | T×8 | T×8 | T×8 | T×8 | T×3 | T×1 |
| # filters |  | 8 |  | 8 | 8 | 8 | 8 | 8 | 3 | 1 |
| filter size |  | 3 |  | 3 | 3 | 3 | 3 | 3 | 1 | 1 |
| dilation |  |  |  |  | 2 | 4 | 8 | 16 |  |  |
| activation |  | ReLU |  | ReLU | ReLU | ReLU | ReLU | ReLU | ReLU | Linear |

**Table 2 - 3D Network. The input signal (a 256x256 pixel wide video of length T) is processed on the single-pixel level by the 21 first layers of the 1D convolutional network (Table 1). Pixel-wise 1D convolutions are implemented by 3D convolutional layers (Cv1 … Cv21) with filter size 1 in the spatial dimensions. The final 3D convolutional layer (Cv22) is effectively a fully connected layer linking all activations in all pixels at a given timestep to a single output neuron. A 20% dropout layer is applied to the input, and another one to the final activation map to avoid overfitting. The output is a univariate time series of length T. act.: activation; ReLU: Rectified linear unit. The final networks are available as Data File 2, Data File 3 and Data File 4.**

|  | Inp | Drop | Cv1 | … | Cv16 | Cv17 | Cv18 | Cv19 | Cv20 | Cv21 | Drop | Cv22 |
|---|---|---|---|---|---|---|---|---|---|---|---|---|
| output dim | 256×256×T×1 | 256×256×T×1 | 256×256×T×8 | | 256×256×T×8 | 256×256×T×8 | 256×256×T×8 | 256×256×T×8 | 256×256×T×8 | 256×256×T×3 | 256×256×T×3 | 1×1×T×1 |
| # filters |  |  | 8 |  | 8 | 8 | 8 | 8 | 8 | 3 |  | 1 |
| filter size |  |  | 1×1×3 |  | 1×1×3 | 1×1×3 | 1×1×3 | 1×1×3 | 1×1×3 | 1×1×1 |  | 256×256×1 |
| dropout |  | 20% |  |  |  |  |  |  |  |  | 20% |  |
| dilation |  |  |  |  |  | 2 | 4 | 8 | 16 |  |  |  |
| act. |  |  | ReLU |  | ReLU | ReLU | ReLU | ReLU | ReLU | ReLU |  | Linear |